# PRELIMINARY DESIGN OF CONTINUOUS WAVE LOW-LEVEL RF SYSTEMS FOR S³FEL*

Z. Y. Zhang[1], J. F. Zhu[1†], H. L. Ding[2‡], J. W. Han[1], W. Li[1], J. Y. Yang[2], W. Q. Zhang[2]
[1]Institute of Advanced Light Source Facilities, Shenzhen, China
[2]Dalian Institute of Chemical Physics, Chinese Academy of Sciences, Dalian, China

*Abstract*

In the Shenzhen Superconducting Soft X-ray Free Electron Laser (S³FEL), Continuous Wave (CW) Low-Level Radio Frequency (LLRF) systems perform critical functions including adjusting the power coupling of accelerator cavities, regulating the amplitude and phase of the RF field, and maintaining the resonance frequency and phase of the cavities. These functions are essential to ensure the electron beam operates at the accelerating phase. Within S³FEL, each superconducting cavity is driven by a solid-state amplifier (SSA), with each SSA paired with a dedicated LLRF system. Based on the distinct acceleration cavities employed, the CW LLRF systems for S³FEL are categorized into four types: 1. Primary accelerator LLRF systems (superconducting, 1.3 GHz; quantity: 168), 2. Harmonic cavity LLRF systems (superconducting, 3.9 GHz; quantity: 16), 3. VHF electron gun LLRF systems (room temperature, 216 MHz; quantity: 4), 4. Buncher LLRF systems (room temperature, 1.3 GHz; quantity: 2). These four LLRF system categories exhibit differing requirements for RF field and acceleration cavity control. This report presents the preliminary design schemes for these four types of CW LLRF systems.

## INTRODUCTION

Free-electron lasers (FEL) [1,2] developed in the 1970s. With the wide applications in photochemistry, materials science, biology and other fields. Developed countries such as the United States, Japan, and Germany have actively started deploying FEL facilities. To obtain high-quality electron beams, it is essential to ensure strict stability in cavity voltage and phase, which low-level radio frequency (LLRF) control of the cavities is necessary. The buncher, VHF electron gun [3] and superconducting cavity (1.3 GHz, 3.9 GHz) has different structures and physical characteristics, so the distinct functions of LLRF need be designed accordingly. This paper primarily introduces the physical performance targets that these types of cavities need to achieve, the LLRF functions designed based on these physical performance targets. After completing the LLRF design, the LLRF functional verification was conducted at the Dalian Advanced Light Source (DALS) and then electron beam transmission experiment at 1MHz was completed. According to the LLRF verification outcomes, the shortcomings need to improve to get better performance for preparing for the Shenzhen S³FEL project.

## LLRF FOR BUNCHER

The buncher use velocity modulation technology to enable electrons operating at the zero-phase point. When electrons that are ahead of the synchronous phase arrive at the bunching cavity, they enter a decelerating phase, whereas electrons that lag behind the synchronous phase and arrive at the cavity experience an accelerating phase. Consequently, electrons near the zero phase are drawn closer to the synchronous electrons, achieving a bunching effect. The physical targets required for the buncher are outlined in the table 1.

Table 1: Physical Targets of Buncher

| Parameter | Value | Unit |
|---|---|---|
| Operating Frequency | 1300 | MHz |
| Detuning | 0~500 | Hz |
| Cavity Voltage | 340 | kV |
| Amplitude Stability | 0.02 | % |
| Phase Stability | 0.02 | ° |

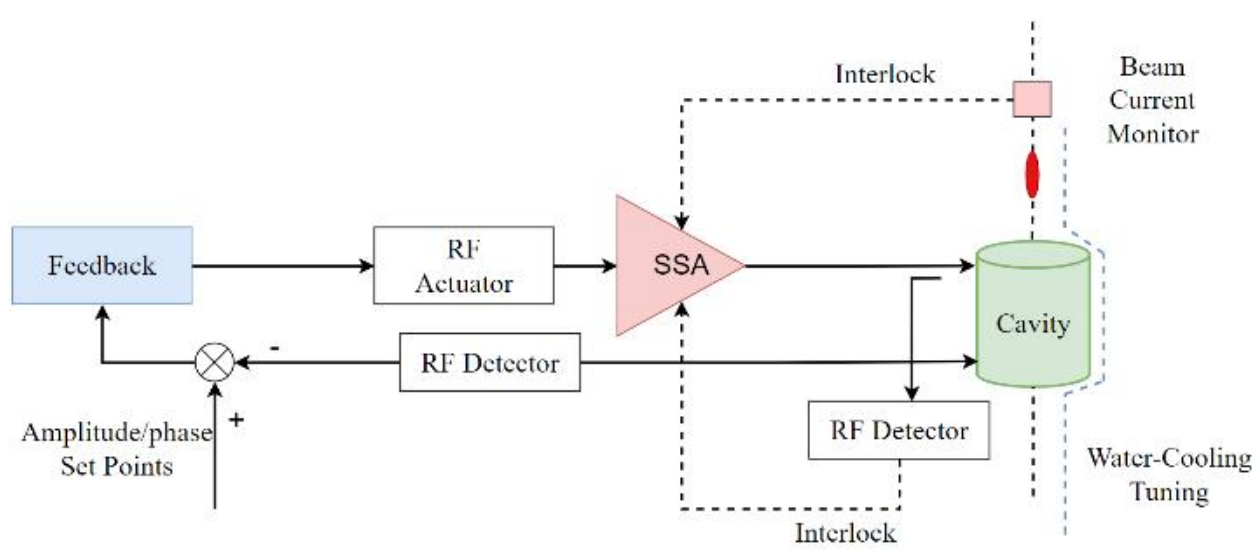

Figure 1: The LLRF control logic for buncher.

The designed half-bandwidth of the buncher is 65 kHz. Detuning caused by external disturbances and noise has a relatively minor impact on the amplitude and phase of the cavity voltage in the buncher. Therefore, tuning control of the buncher can be achieved using relatively slow water-cooling tuning methods, while the LLRF primarily implements RF control to stabilize the amplitude and phase of the cavity voltage. The room-temperature cavity is a traveling-wave cavity, and excessive reflected power can lead to worse vacuum and waveguide window arcing. Hence, it is essential to monitor the reflected power and design an interlock system to shut down the power supply of SSA when necessary. The LLRF control logic diagram applied to the buncher is shown in the Fig. 1.

* Work supported by the National Natural Science Foundation of China (Grant No. 12405221).
† zhujinfu@mail.iasf.ac.cn
‡ dinghongli@dicp.ac.cn

When the designed LLRF is applied to the buncher of the DALS, the buncher can achieve a cavity voltage of 340 kV at a reference frequency of 1.3 GHz, with closed loop amplitude stability of 0.0076% and closed loop phase stability of 0.0067°. These performance meets the physically required targets. The time-domain curves of amplitude noise and phase noise are shown in the Fig. 2.

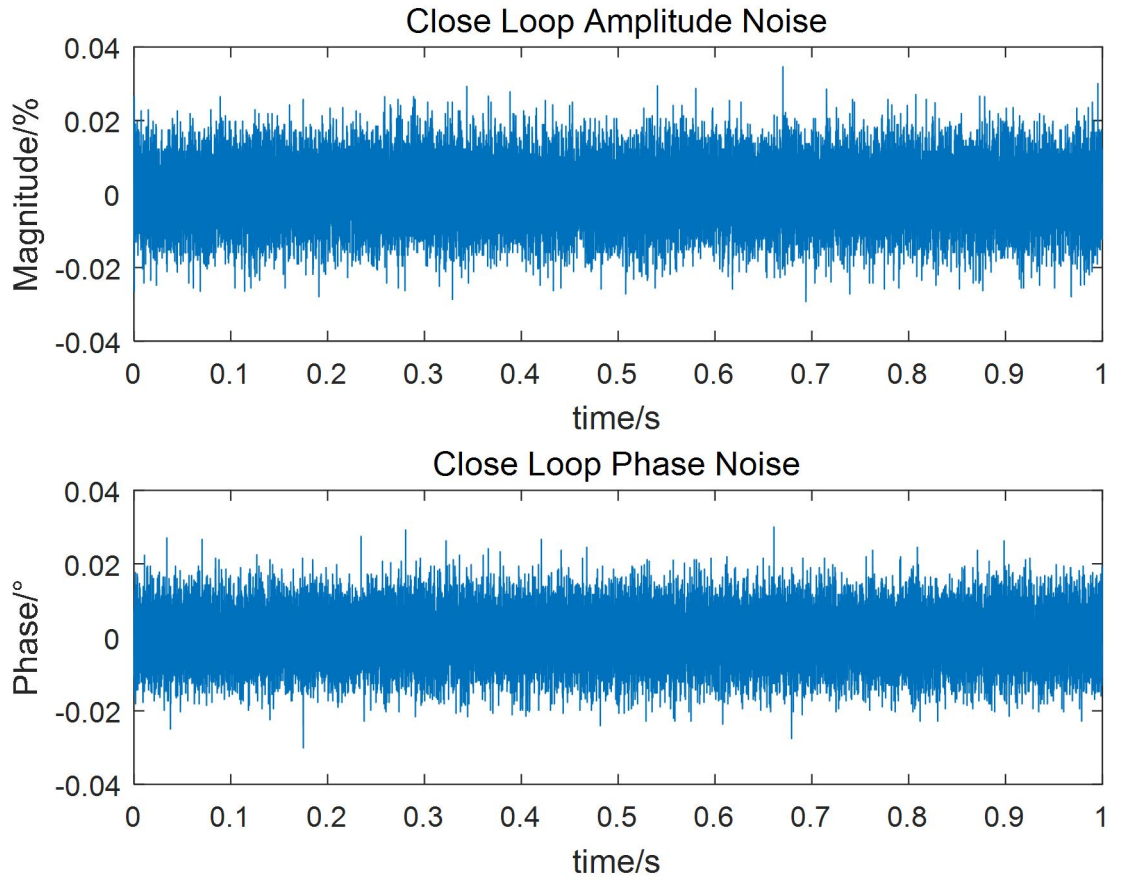

Figure 2: Close loop amplitude noise and phase noise.

## LLRF FOR VHF ELECTRON GUN

Electron guns that can operate under continuous-wave conditions include VHF electron guns, Radio Frequency (RF) superconducting electron guns, Direct Current (DC) high-voltage electron guns, and DC high-voltage-RF superconducting electron guns. The VHF electron gun has advantages of relatively high electric field intensity at the photocathode and a relatively large physical size that facilitates heat dissipation, so it is enable for long-time operation under continuous wave at room temperature. The physical targets required for the VHF electron guns are outlined in the table 2.

Table 2: Physical Targets of VHF Electron Gun

| Parameter | Value | Unit |
|---|---|---|
| Operating Frequency | 216.67 | MHz |
| Detuning | 0~100 | Hz |
| Cavity Voltage | 750 | kV |
| Amplitude Stability | 0.02 | % |
| Phase Stability | 0.02 | ° |

Unlike the buncher, the electron gun has a half-bandwidth of only 7 kHz, and detuning caused by cavity temperature and environmental vibrations can affect the cavity voltage and phase. Therefore, to meet the requirements for tuning frequency accuracy, VHF electron gun employs four stepper motors for tuning. Additionally, VHF electron gun utilizes two SSA inputs. To ensure that the power and phase remain balanced between two SSA, taking the difference of two SSA output as the feedback control's input to change one SSA output to following the other SSA output. The LLRF control logic diagram applied to the VHF electron gun is shown in the Fig. 3.

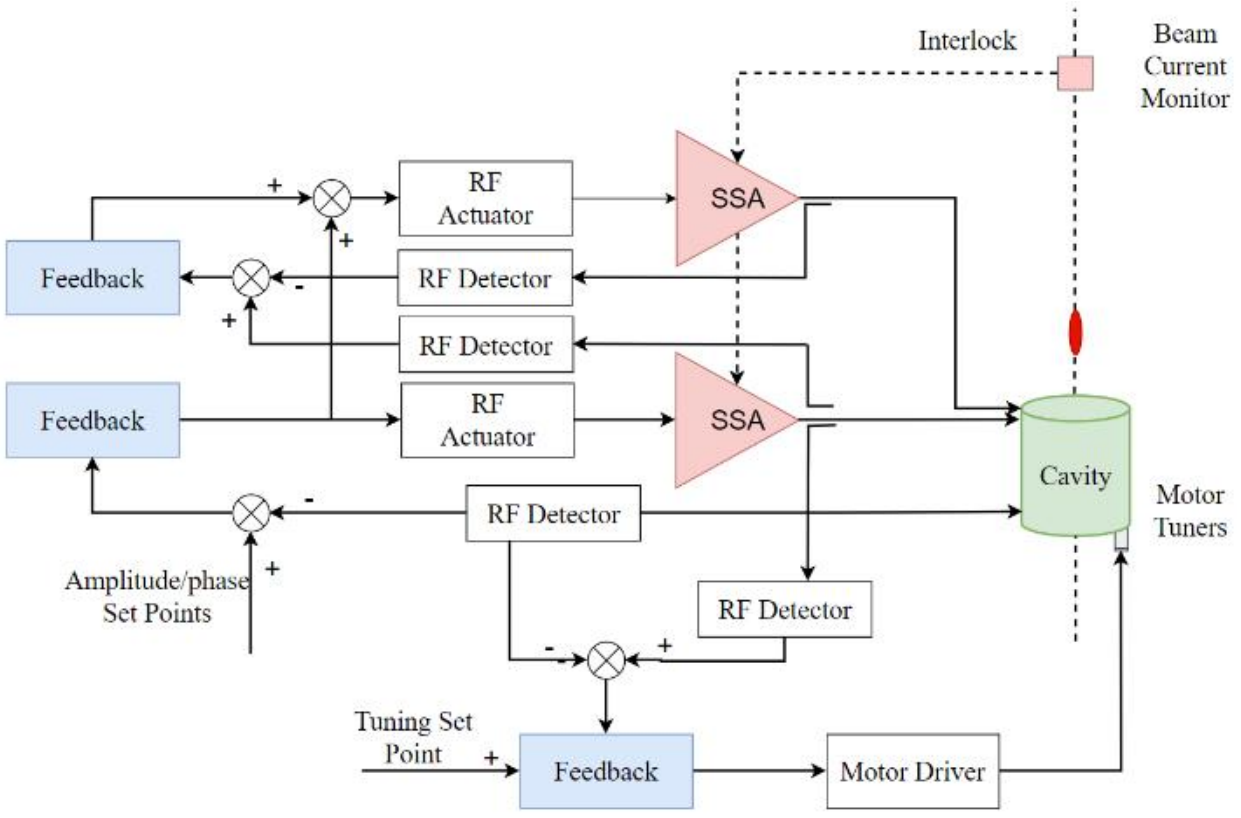

Figure 3: The LLRF control logic for VHF electron gun.

For testing the designed LLRF performance, it is applied to the VHF electron gun of the DALS, the VHF electron gun can achieve a cavity voltage of 750 kV at a reference frequency of 216.67 MHz, with closed loop amplitude stability of 0.0086% and closed loop phase stability of 0.0055°. These performance meets the physically required targets. The time-domain curves of amplitude noise and phase noise are shown in the Fig. 4.

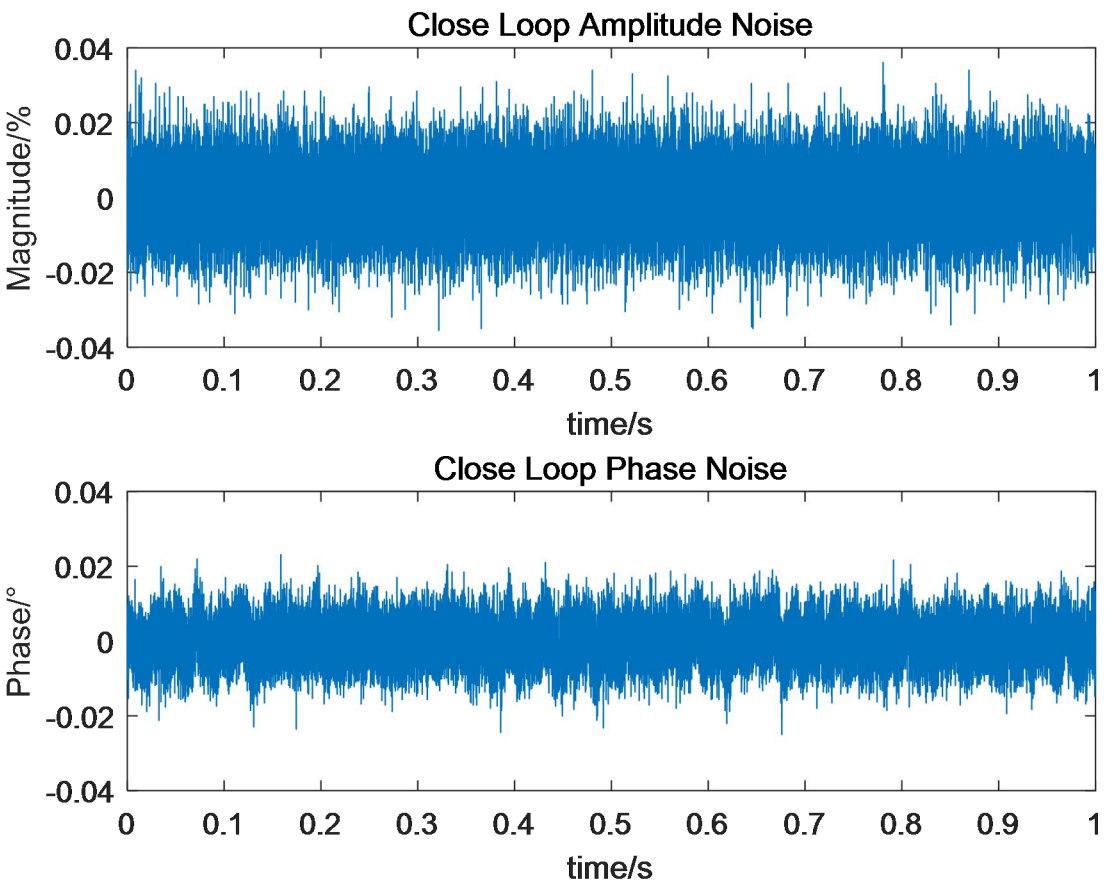

Figure 4: Close loop amplitude noise and phase noise.

## LLRF FOR SUPERCONDUCTING CAVITY

The role of the superconducting cavity is to accelerate the electron beam and enhance its energy. The superconducting cavities include primary accelerator (1.3 GHz) and Harmonic cavity (3.9 GHz) types, which differ only in frequency while having same control loops and interlock systems. The physical targets required for the superconducting cavity are outlined in the table 3.

Compared with normal-conducting cavities, superconducting cavities have a half-bandwidth of 30 Hz ($Q_L = 2.0 \times 10^7$) around, Some cavity half-bandwidths can reach up to 15 Hz ($Q_L = 4.3 \times 10^7$). Piezeo [4,5] as a tuning actuator of high precision needs to be used for superconducting cavities. The control method employs a combination of feedback and feedforward control. This is because delays exist in the tuning loop, which has a bad affect on the feedback control performance for high frequency detuning. Therefore, feedforward control appling the state machine approach is adopted to suppress high frequency detuning. The LLRF control logic diagram applied to the superconducting cavity is shown in the Fig. 5.

Table 3: Physical Targets of Superconducting Cavity

| Parameter | Value | Unit |
|---|---|---|
| Operating Frequency | 1300 | MHz |
| Detuning | 0~2 | Hz |
| Cavity Voltage | 16 | MV |
| Amplitude Stability | 0.02 | % |
| Phase Stability | 0.02 | ° |

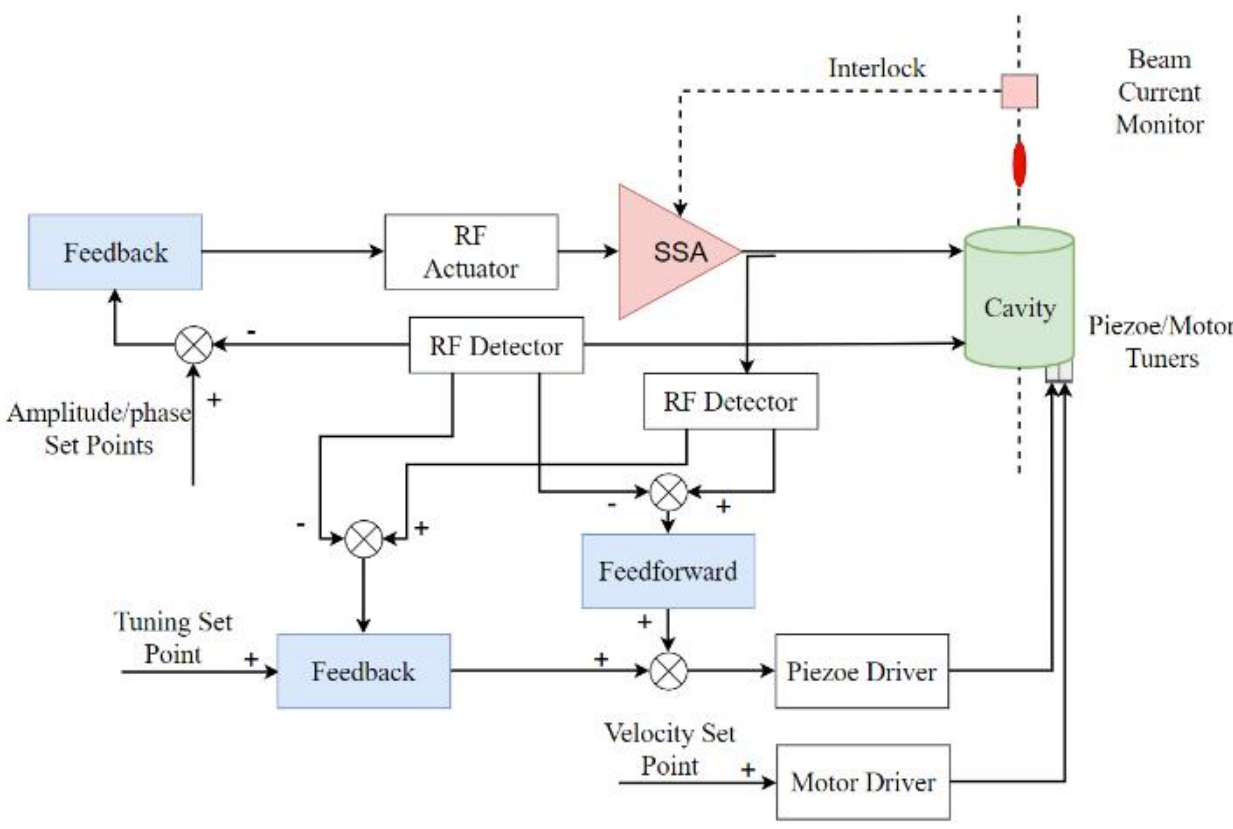

Figure 5: The LLRF control logic for superconducting cavity.

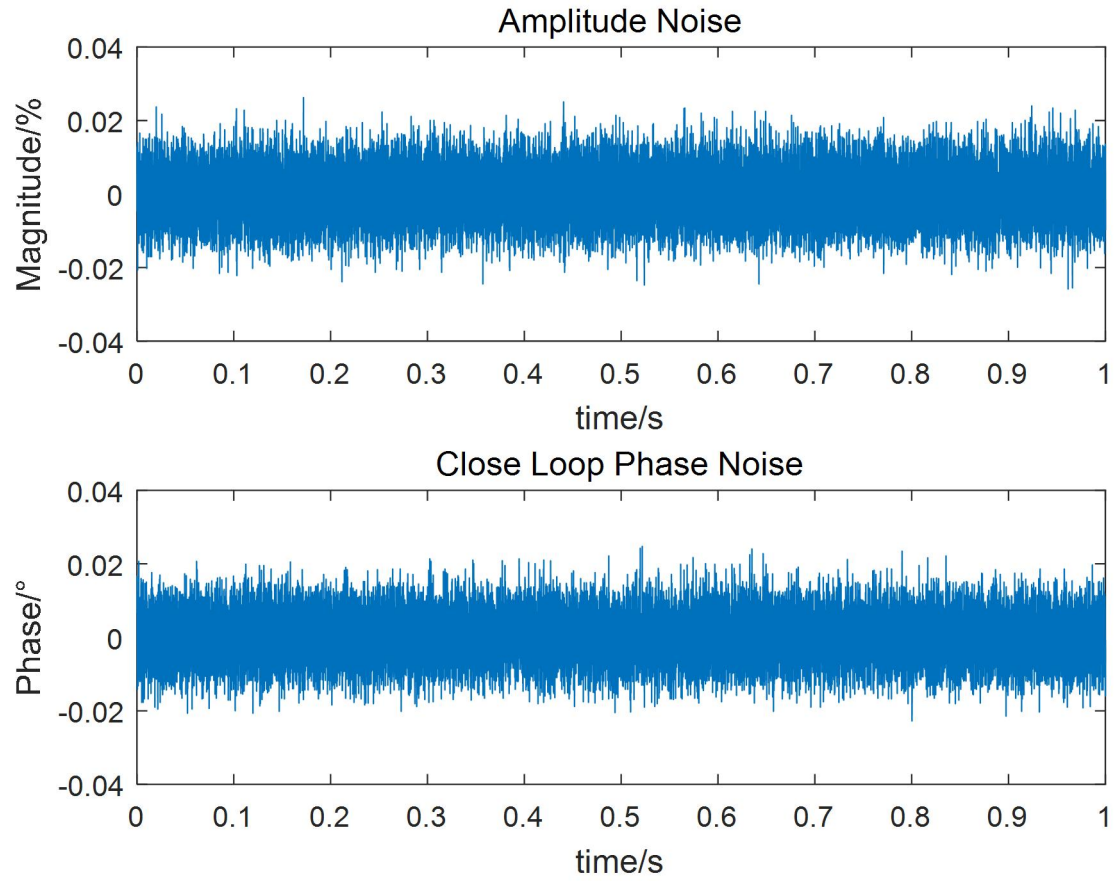

Figure 6: Close loop amplitude noise and phase noise.

Applying the designed LLRF to the superconducting cavity of the DALS, the superconducting cavity can achieve a cavity voltage of 16 MV at a reference frequency of 1300 MHz, with closed loop amplitude stability of 0.0066% and closed loop phase stability of 0.0061°. These performance meets the physically required targets. The time-domain curves of amplitude noise and phase noise are shown in the Fig. 6.

## CONCLUSION

This paper introduced four types of LLRF systems for Shenzhen S[3]FEL project, these four LLRF systems have been verified at the DALS platform and meet the specified physical requirements. Besides, there are shortcomings worthy of improvement, such as further enhancing the stability of amplitude and phase for preparation for the Shenzhen project.